\documentclass{aa}  % for referee version [referee], change figure* (2-column) to \textwidth

\usepackage{graphicx}
\usepackage{txfonts}
\usepackage{xcolor}
\usepackage{amssymb}
\usepackage[]{hyperref}
\begin{document} 

\title{Hanle rotation signatures in Sr~{\sc i} 4607\,\AA}
\titlerunning{Hanle rotation in Sr~{\sc i} 4607\,\AA}

   \author{F. Zeuner \inst{1}
          \and
          L. Belluzzi \inst{1}$^,$\inst{2}$^,$\inst{3}
           \and
          N. Guerreiro \inst{1}
          \and
           R. Ramelli \inst{1}
          \and
          M. Bianda
          \inst{1}$^,$\inst{3}
          }

   \institute{
   $^1$ IRSOL Istituto Ricerche Solari ``Aldo e Cele Dacco'', Università della Svizzera italiana, 6605 Locarno, Switzerland\\
   $^2$ Leibniz-Institut für Sonnenphysik (KIS), 79104 Freiburg i. Br., Germany\\
   $^3$ Euler Institute, Università della Svizzera italiana (USI), 6900 Lugano, Switzerland
   }
   \date{Submitted: February 17, 2022, Accepted: April 4, 2022}

 \abstract 
    {
    Measuring small-scale magnetic fields and constraining their role in energy transport and dynamics in the solar atmosphere are crucial, albeit challenging, tasks in solar physics. 
    To this aim, observations of scattering polarization and the Hanle effect in various spectral lines are increasingly used to complement traditional magnetic field determination techniques.
   }
   {
    One of the strongest scattering polarization signals in the photosphere is measured in the Sr~{\sc i} line at 4607.3\,\AA\ when observed close to the solar limb. 
    Here, we present the first observational evidence of Hanle rotation in the linearly polarized spectrum of this line at several limb distances.
   }
   {
    We used the Zurich IMaging POLarimeter, ZIMPOL at the IRSOL observatory, with exceptionally good seeing conditions and long integration times. 
    We combined the fast-modulating polarimeter with a slow modulator installed in front of the telescope. 
    This combination allows for a high level of precision and unprecedented accuracy in the measurement of spectropolarimetric data.   }
   {
    Fixing the reference direction for positive Stokes $Q$ parallel to the limb, we detected singly peaked $U/I$ signals well above the noise level. We can exclude any instrumental origins for such $U/I$ signals.
    These signatures are exclusively found in the Sr~{\sc i} line, but not in the adjoining Fe~{\sc i} line, therefore eliminating the Zeeman effect as the mechanism responsible for their appearance. However, we find a clear spatial correlation between the circular polarization produced by the Zeeman effect and the $U/I$ amplitudes.
    This suggests that the detected $U/I$ signals are the signatures of Hanle rotation caused by a spatially resolved magnetic field. 
   }
   {
   A novel measurement technique allows for determining the absolute level of polarization with unprecedented precision. Using this technique, high-precision spectropolarimetric observations reveal, for the first time, unambiguous $U/I$ signals attributed to Hanle rotation in the Sr~{\sc i} line.
   }


   \maketitle

%-------------------------------------------------------------------
%-------------------------------------------------------------------
\section{Introduction}
\label{sec:introduction}
%-------------------------------------------------------------------

Since its discovery more than four decades ago in the spectrum close to the solar limb \citep{Wiehr1978,Stenflo1980}, the linear polarization arising from scattering processes in the Sr~{\sc i} line at 4607.3\,\AA\ has drawn substantial attention.
This strong scattering polarization signal is unrivaled within all photospheric spectral lines, reaching amplitudes above 1\% close to the limb (see, e.g., \citealp{Stenflo1997a} and \citealp{Gandorfer2002}). 
Its sensitivity to the Hanle effect \citep{Hanle1924} has been exploited to provide valuable insights into weak photospheric magnetic fields with complex topologies \citep[e.g.,][]{Stenflo1982, Faurobert-Scholl1993,TrujilloBueno2004,Bellot2019}.
These magnetic fields are especially interesting in regions where a local solar dynamo might operate \citep[][]{Petrovay1993, Vogler2005, Rempel2014}, whereby questions regarding the mere existence and possible energetic coupling mechanisms to the outer layers of the atmosphere are still unsettled.\par
The Hanle effect leaves two characteristic traces in the core of spectral lines that are linearly polarized under scattering processes. 
First, it modifies (generally reducing) the linear polarization degree, such a variation depending on the unsigned magnetic field strength.
Therefore, even turbulent magnetic field topologies can be detected by suitably interpreting this signature of the Hanle effect.
Second, if the magnetic field structure within a resolution element is deterministic, namely, it is characterized by a net magnetic flux unequal to zero, the linear polarization plane is additionally rotated \citep{LandiDeglInnocenti2004}. 
Hanle rotation signatures (i.e., appearance of Stokes $U$ signals) have been clearly observed in the He~{\sc i} lines at 10830 and 5876\,{\AA} in solar prominences and filaments. 
Such signals have been used to infer information on magnetic fields permeating these plasma structures 
\citep[see most recent works by][]{Orozco_Su_rez_2014,2020A&A...640A..71K}. 
Similar signals have also been observed in strong chromospheric lines such as Sr~{\sc ii} at 4078\,\AA\ \citep{Biandaetal1998} and
Ca~{\sc i} at 4227\,\AA\ \citep[e.g.,][]{Biandaetal1999,Stenflo2003b,Anusha2011,2020A&A...641A..63C}. 
To the best of our knowledge, Hanle rotation signatures in the Sr~{\sc i} line have once been reported in the review by \citet{Bellot2019}, and are referred to as ``not common''. 
These Stokes $U$ signals in Sr~{\sc i}, due to a deterministic magnetic field, should not be confused with Stokes $U$ theoretically predicted \citep{DelPinoAleman2018} and observationally detected \citep{Zeuner2020} at the center of the solar disk, which is predominantly caused by the axial symmetry break of the radiation field due to local inhomogeneities of the photosphere. 
We recall that the Hanle critical field \citep[i.e., the magnetic field strength at which the sensitivity to the Hanle effect is maximum; see][]{jtb2001} for the Sr~{\sc i} 4607\,{\AA} line, in the absence of collisions, is about 20\,G. This value becomes higher when elastic collisions, which have a significant impact during the formation of this photospheric line, are taken into account \citep{DelPinoAleman2018}.\par

Essential prerequisites to detect Hanle rotation are high-precision and high-sensitivity spectropolarimetric observations. 
The ZIMPOL system \citep{Ramelli2010} has proven invaluable for high-sensitivity scattering polarization observations \citep[e.g.,][]{Stenflo1996, Stenflo1997, Stenflo2003b, Bianda2011, AlsinaBallester2021} due to its high modulation frequencies of 1\,kHz or more. 
These modulation frequencies guarantee that a main error source in ground-based polarimetric observations, such as polarization cross-talk induced by the seeing by Earth's atmosphere, is eliminated and polarimetric precision at the photon shot-noise limit can be reached \citep{Stenflo2013}. 
However, when ZIMPOL is attached to the IRSOL telescope system, the accuracy of the observed Stokes parameters may suffer from additional instrumental polarization, such as artificial offsets due to for example telescope polarization, or instrumental fringes. 
The linear polarization Stokes parameters may additionally suffer from Stokes $V$ cross-talk.
These effects complicate a clear polarimetric detection in cases where the polarization's zero level is crucial. 
For Hanle rotation signatures, where the sign of Stokes $U$ depends on the direction of the magnetic field and therefore is expected to be either negative or positive, such a precise zero level determination is vital. \par
A technique for a zero level determination is already successfully implemented at the IRSOL observatory. 
It is based on an additional slow modulation in front of the telescope (Zeuner et al., in prep.). The advantages of this technique have been used to measure the continuum polarization level with respect to an unprecedentedly precise zero level in a broad wavelength range (Berdyugina et al., in prep.) in contrast to previous efforts \citep[e.g.,][]{1972A&A....19..287L,1974A&A....31..179M,1975A&A....38..303W}. 
Furthermore, the benefits of this technique for large aperture solar telescopes are studied within the SOLARNET project, in preparation for the 4\,m European Solar Telescope.\par

In the present paper, we show observational evidence of Hanle rotation signatures in a spectropolarimetric observation of the Sr~{\sc i} 4607.3\,\AA\ line carried out at the IRSOL observatory with the ZIMPOL system. 
The polarimeter is used in combination with a slow modulator in front of the telescope to provide high-precision observations with high zero level accuracy. 
We focus in the investigation of the linear polarization signals whose orientation deviates from the solar limb tangent and their possible correlations with the line-of-sight magnetic field, which is evidenced by circular polarization produced by the Zeeman effect.

%-------------------------------------------------------------------
\section{Observations}
\label{sec:observations}
%-------------------------------------------------------------------

\begin{figure*}
\centering
\includegraphics[width=17cm]{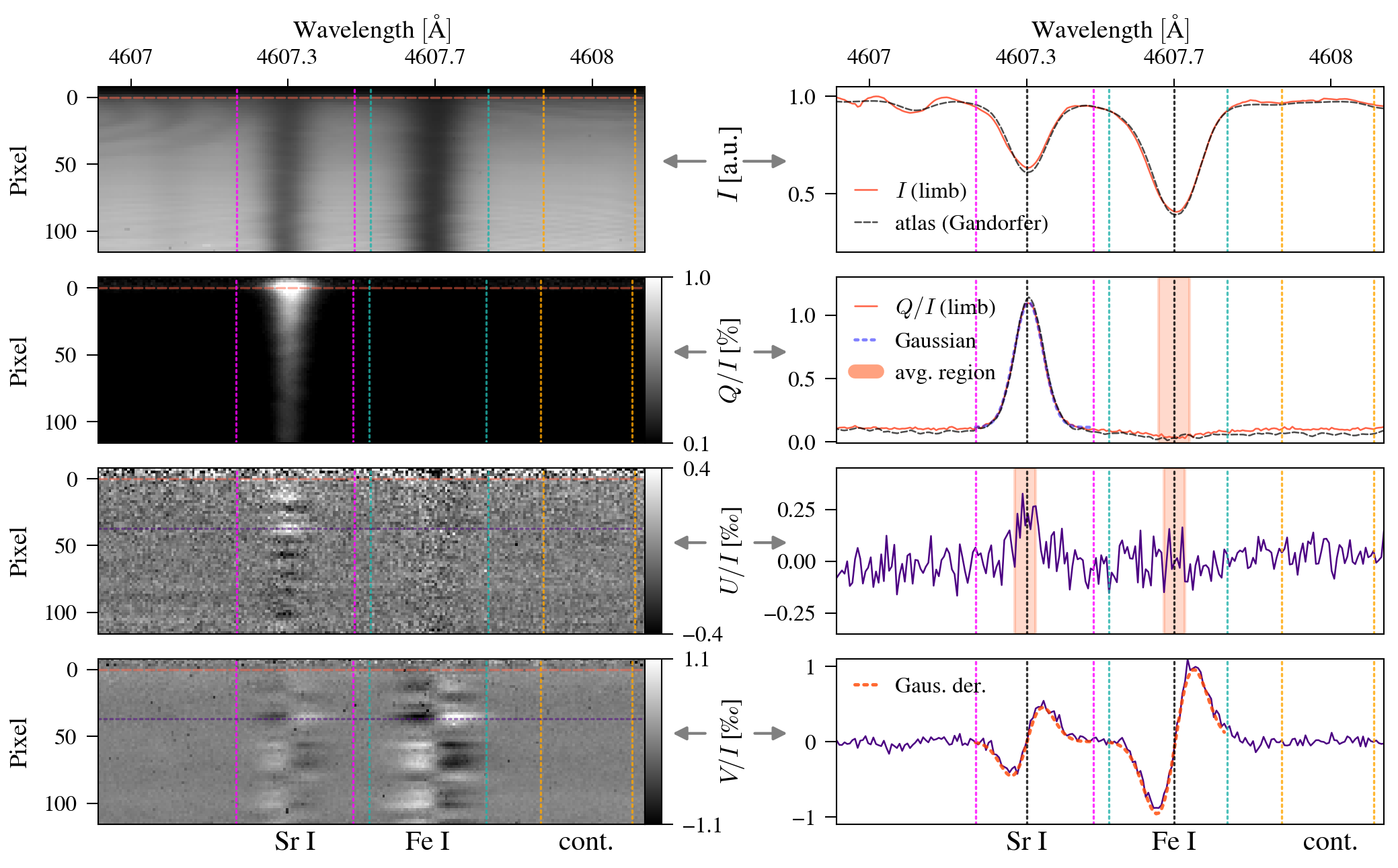} %17cm
 \caption{
    ZIMPOL Stokes images containing different limb distances (\textit{left}) and example spectra at specific locations (\textit{right}).
    \textit{Four left panels:} $I$, $Q/I$, $U/I$, and $V/I$ (\textit{top to bottom}, respectively) Stokes images after the data reduction. 
    The CLV is along the spatial pixels of the camera. 
    Three spectral regions are identified in the images: Sr~{\sc i}, Fe~{\sc i,} and continuum, indicated by the colored dotted vertical lines (Sr~{\sc i} = magenta, Fe~{\sc i} = cyan, continuum = orange). 
    The limb position is indicated by the horizontal orange-red dashed line.
    \textit{Right panels} show example spectra: Stokes $I$ and $Q/I$ close to the limb \citep[together with the spectra given in the Second Solar Spectrum atlas by][]{Gandorfer2002}, Stokes $U/I$ and $V/I$ in a magnetically active region (position indicated by the purple lines in the left images). 
    The $Q/I$ in the Sr~{\sc i} region is fitted with a Gaussian, while $V/I$ in Sr~{\sc i} and Fe~{\sc i} is fitted with a Gaussian derivative. 
    Regions that are averaged for further data analysis are indicated by the light orange bands. 
}
\label{fig:obs_images}
\end{figure*}

Our observation was performed with the Zurich IMaging POLarimeter, ZIMPOL-3 
\citep[hereafter ZIMPOL; see][]{Ramelli2010}, at the 45\,cm aperture IRSOL Gregory Coud\'e Telescope in Locarno, Switzerland, together with a Czerny Turner spectrograph. 
An inteference prefilter at 4607\,\AA,\ with a width of 300\,\AA\, and 65\% transmission was used to avoid overlapping spectrograph orders.
The slit width of the spectrograph was 60\,µm (corresponding to 0.5\,{\ensuremath{^{\prime\prime}}}) and the 
spectral resolution per pixel is 5.7\,m\AA. 
We carried out our observations on August 11, 2021 at 14:36\,UT for 73\,minutes with an integration time of 0.6\,s per frame. 
In total, 2800 frames were recorded. 
The seeing conditions were exceptionally good. 
The observed region was the solar west limb, showing no signs of magnetic activity (see low circular polarization values in Figure~\ref{fig:obs_images}). 
The slit was oriented perpendicular to the limb, thereby covering various limb positions at the same time. \par

The ZIMPOL camera is combined with a photo elastic modulator \citep[PEM, the implementation details can be found in][]{Gandorfer1997}, running at 42\,kHz, followed by an achromatic polarizing beamsplitter, both placed directly after the polarization calibration unit (PCU) but before the slit. 
The PCU is mounted after two folding mirrors, thus some cross-talk is unavoidable and needs to be accounted for in the data reduction.\par
An image de-rotator based on three mirrors is used to fix the orientation of the solar limb with respect to the slit.
The image de-rotator is placed after the modulator, both rotating synchronously, but the de-rotator at half angular velocity speed to keep the polarimetric reference system fixed with respect to the slit. 
The angle error between the slit and limb is estimated to be better than half a degree, potentially resulting in some $Q$$\leftrightarrow$$U$ cross-talk. 
The position on the Sun is controlled by the Primary Image Guiding system \citep{1998SoPh..182..247K,2011AN....332..502K} which, under stable weather conditions, has a relative error of about 1.5\,{\ensuremath{^{\prime\prime}}}. 
To further stabilize the limb position with respect to the slit, the limb distance on the slit-jaw camera image is calculated two times per second during the observation. A glass plate, with a rotation axis parallel to the limb, is used to correct for the residual shifts. \par
Next, we briefly describe the technical setup and the novel measurement technique based on the slow modulation. For more details on the reduction and the theoretical description, we refer to the upcoming manuscript that is to be published by Zeuner et al. (in prep.).
To perform the slow modulation, a polymer-based zero-order retarder film that covers the complete telescope aperture is mounted in front of the telescope on the Telescope Calibration Unit (TCU) --- a large, but light, ball bearing moved by a step motor with high precision. Initially, the fast axis is oriented along the observed solar limb.  
The accuracy of the fast axis orientation is about 1$^\circ$ that can result in $Q$$\leftrightarrow$$U$ cross-talk. 
The film used for the measurements in this paper is a half-wave retarder at 3900\,\AA. 
For a wavelength of 4607\,\AA\ the retardance of this film is 0.39 (in units of wave shifts)\footnote{Due to the large dimensions of the film, the retardance could only be characterized for a small part of it.}. The implication of this reduced retardance is that the polarimetric efficiency \citep[as in][]{delToroIniesta2000} of measuring the linear polarization for Sr~{\sc i} is decreased to a factor of 0.65 (an efficiency of one is achieved at the design wavelength of 3900\,\AA).
Every 20 frames, the film is rotated by 22.5$^\circ$, thereby slowly modulating the sign of the incoming polarization. 
Eight angle positions of the TCU are needed for a complete modulation cycle to measure the full Stokes vector \citep[rendering obsolete a change of the PEM orientation that was needed in previous measurements with the same setup, e.g.,][]{Bianda2018}. 
As all other effects compromising the measurement are introduced after the TCU (e.g., cross-talk $V$$\leftrightarrow$$\{Q,U\}$ and offsets by the two folding mirrors inside the telescope), a simple subtraction removes all defects that are temporally stable during the measurement. 
The temporal stability of these effects over the course of one hour is given by the slowly varying telescope matrix \citep[i.e., the relative orientation of the two folding mirrors of the IRSOL telescope only changes with the declination, see][]{Ramelli2006}. 
The subtraction is applied during the standard data reduction (see Sect.~\ref{sec:dare_clv}).\par

The polarimetric calibration, the dark image and the flat-field image (obtained by moving the telescope randomly around disc center) were all recorded within a few minutes before and after the observations. Stokes $+Q$ is defined  parallel to the limb. 
Additionally, we calibrated the pixel scale of the ZIMPOL camera on the day of the observation to (1.29 $\pm$ 0.08)\,{\ensuremath{^{\prime\prime}}}\,pixel$^{-1}$. 

\subsection{Data reduction}
\label{sec:dare_clv}

The acquired ZIMPOL Stokes data were corrected for dark images, polarimetrically calibrated by applying the demodulation matrix, and flat-field corrected. 
The flat-field only applies to the Stokes $I$ and is used to reduce the fringes in the intensity images, otherwise our analysis is independent of this step, as the fringes are unpolarized. 
Since the data collection spans more than one hour, the data were split and the temporally closest flat-field, dark images, and demodulation matrices were used.
The flat-field correction could only be performed partially due to the long observing time, resulting in residual fringes in Stokes $I$.
The modulation matrix changes only marginally, and we did not find a significant change in the polarization amplitude when splitting the data or demodulate the full set with the first demodulation matrix. After demodulation, the Stokes images were combined according to the slow modulation to correct for most instrumental errors. 
In total, 35 corrected full Stokes images have been recorded during this observation. \par

However, the $Q/I$$\leftrightarrow$$U/I$ cross-talk induced by the telescope mirrors or by an imperfectly aligned TCU film cannot be corrected using the slow modulation technique.  
Usually, this cross-talk is low, on the order of 1-2\%, and is not significant for measurements with low signal-to-noise ratios (S/N) in Stokes $U/I$. 
Here, our S/N is significantly higher than usual, and a small $Q/I$$\rightarrow$$U/I$ cross-talk is  noticeable in all wavelengths, but naturally most prominently in the Sr~{\sc i} line. 
We corrected this cross-talk with an ad-hoc subtraction of 1.5\% of $Q/I$ from $U/I$. 
The value of 1.5\% was found by minimizing the cross-talk $Q/I$$\rightarrow$$U/I$ in the continuum region of $U/I$, where only noise is expected. 
We stress that the subtraction mainly affects the amplitudes of $U/I$ in the Sr~{\sc i} line where $Q/I>0.5\%$ (i.e., very close to the limb), because for smaller $Q/I$ signals the cross-talk is orders of magnitude smaller than the $U/I$ signals. 
Therefore, the $Q/I$$\rightarrow$$U/I$ cross-talk influences $U/I$ solely for small limb distances. Later, we limited our analysis to larger limb distances, and the final results of the paper are independent of this cross-talk correction step. 
To increase the S/N further, we binned over two spatial pixels. 
We find that the binning process does not degrade significantly the spatial resolution. This likely reflects the impact of a long exposure time allied with seeing conditions that are known to decrease the spatial resolution. \par

\begin{figure}
\centering
\includegraphics[width=\hsize]{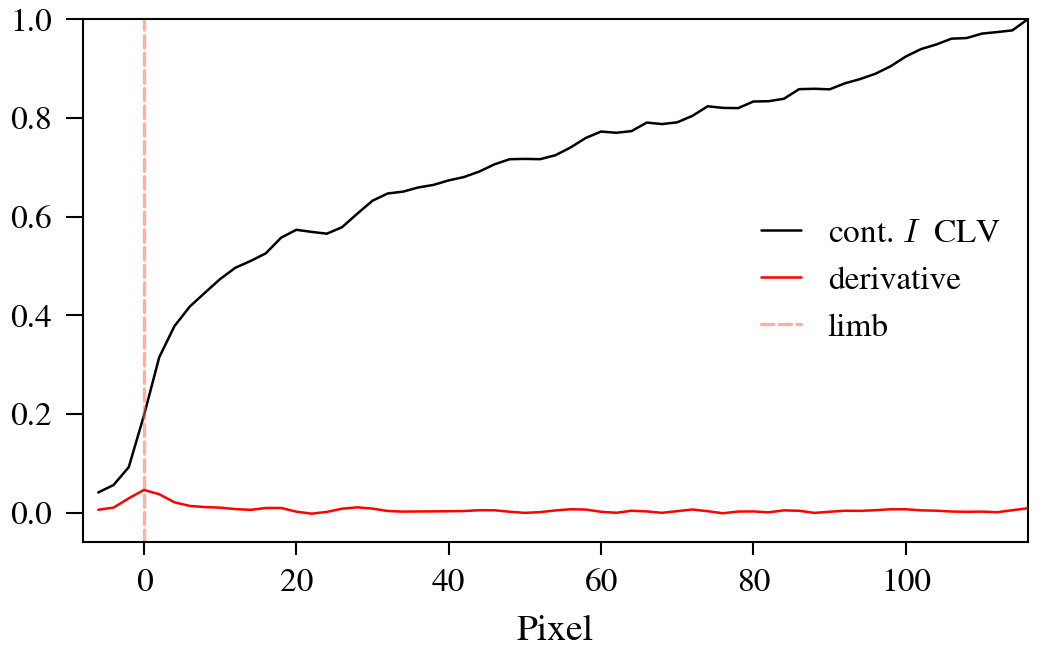}
\caption{
 Center-to-limb variation of the continuum intensity (black line). The limb position (red-dashed vertical line) is determined by the inflection point, i.e., where the first derivative (red line) is at maximum. 
}
\label{fig:limb}
\end{figure}

To find the limb position, Stokes $I$ was averaged across the spectral interval of the continuum (three different spectral regions, indicated with different colors in Fig.~\ref{fig:obs_images}, will be considered in the observed wavelength interval). Then, the inflection point of the center-to-limb variation (CLV) was determined by calculating the maximum of the first derivative. This procedure was repeated for each corrected Stokes $I$ image, see Fig.~\ref{fig:limb} for one example of the intensity CLV and the corresponding derivative. 
The maximum of the derivative is obtained by fitting a second degree polynomial in the proximity of the inflection point. 
We find that the inflection point is shifted less than one pixel between each of the 35 corrected Stokes images. 
This shows the long-term stability of the limb position with respect to the ZIMPOL camera. 
The pixel of the average limb position in all Stokes images is set to zero. 
The final Stokes images averaged over all 35 corrected images are shown in the left panels of Fig.~\ref{fig:obs_images}. With the determined limb position in combination with the ZIMPOL camera pixel scale and the apparent size of the Sun on that day the distance of each pixel from the solar limb is given by $\mu=\cos(\theta)$, where $\theta$ is the heliocentric angle and $\mu=0$ at the solar limb. \par
Since we are not interested in an accurate spectral line profile determination, we did not correct the profiles for spectral stray light contributions, which for the IRSOL spectrograph is typically on the order of 2\%. Stray light contributions from the sky, telescope mirrors, and spectrograph are most crucial very close to the limb. As we later limit our analysis to limb distances $\mu>0.2$, we assume that these contributions do not compromise our analysis in a significant way.

\subsection{Amplitude determination}
\label{sec:amplitude}

In this section, we describe how the $Q/I, U/I$ and $V/I$ amplitudes are determined for further analysis from the corrected Stokes images. We note that the amplitudes are determined separately for each spatial pixel, that is, for each limb distance.
First, the centers of the spectral lines in the intensity spectrum are determined by Gaussian fitting and then used in the averaging process (described in the next paragraph).  \par
For the $Q/I$ profile in the Sr~{\sc i} region (see Fig.~\ref{fig:obs_images} for the definition), the S/N is high. Therefore, a Gaussian (with an offset due to the continuum polarization) was fitted and the amplitude saved for later analysis. 
An example fit is shown in the right $Q/I$ panel of Fig.~\ref{fig:obs_images}. 
In the Fe~{\sc i} region, an average of 12 pixels in spectral dimension symmetrically around the line center position was calculated. Averaging twelve pixels was sufficient to achieve an acceptable S/N (see the $Q/I$ CLV in Fig.~\ref{fig:clvs}).\par

Although the $U/I$ signals are visible in the Stokes $U/I$ images in Sr~{\sc i}, the S/N of the profiles is still too low to fit a Gaussian (an example profile is shown in the right $U/I$ panel of Fig.~\ref{fig:obs_images}). 
Therefore, the contribution of six spectral pixels around the line center is used to calculate an average value. 
The choice of six pixels reflects a compromise between achieving low noise levels, but at the same time avoiding a significant signal reduction. 
In fact, averaging six pixels of a synthetic Gaussian with amplitude one and a standard deviation given by the Sr~{\sc i} line width results in a value of 0.93.  
The regions averaged are indicated in the right panels of Fig.~\ref{fig:obs_images} as light orange bands.\par

The $V/I$ amplitude was determined by fitting the first derivative of a Gaussian, as given by the weak-field-approximation \citep[e.g.,][]{LandiDeglInnocenti2004}. We find that the sign of the red peak
of $V/I$ in Sr~{\sc i} shows some correlation with the sign of $U/I$. 
Therefore, the amplitude of this peak was used in the subsequent analysis. 
The S/N is sufficient to fit all the $V/I$ profiles in Sr~{\sc i} and Fe~{\sc i}. We show examples of the fitted Gaussian derivative in the right $V/I$ panel of Fig.~\ref{fig:obs_images}.

%-------------------------------------------------------------------
\section{Results}
\label{sec:results}
%-------------------------------------------------------------------

\subsection{Center-to-limb variation of polarization amplitudes}
\label{sec:clv_amp}

\begin{figure*}
\centering
\includegraphics[width=17cm]{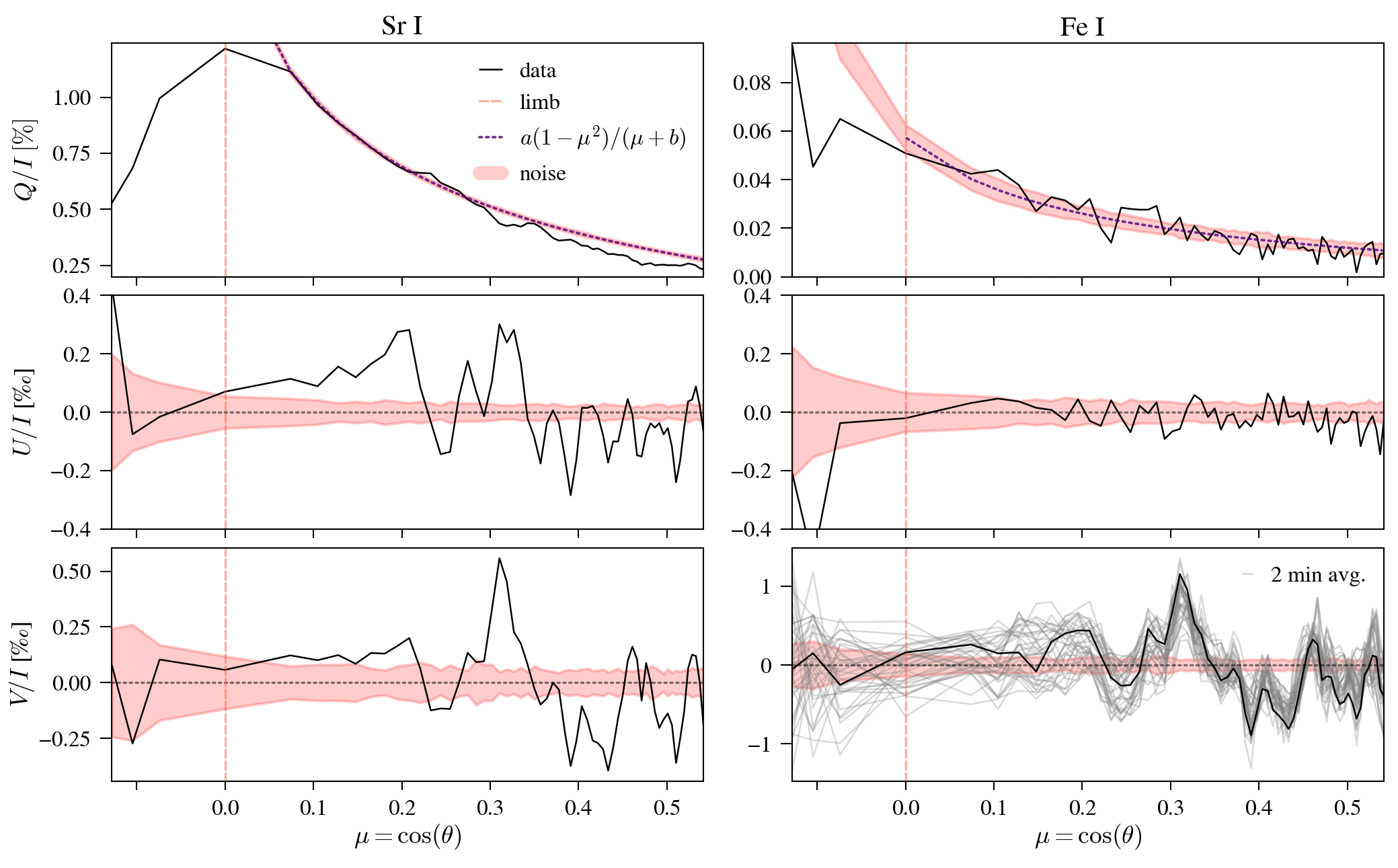} %17cm
 \caption{
    Center-to-limb variations of the $Q/I$, $U/I$, and $V/I$ amplitudes 
    in Sr~{\sc i} (\textit{left}) and Fe~{\sc i} (\textit{right}) shown as black solid lines. 
    The limb position is indicated by the vertical dashed-orange line. 
    The estimated noise level is shaded in red. The CLV of $Q/I$ is fitted with the function presented in the text, shown as a black-dashed line (for $U/I$ and  $V/I$ the black-dashed line is set to zero). 
    At this scale, the estimated noise level band in $Q/I$ of Sr~{\sc i} is very small. 
    For $V/I$ in Fe~{\sc i} we additionally plot with gray lines the red peak averaged over $\sim$2\,min and 15 spectral pixels. 
    See text for details on the parameters $a$ and $b$ for the function (black dotted line) fitted to the $Q/I$ CLV.
}
\label{fig:clvs}
\end{figure*} 

The amplitudes in $Q/I$, $U/I,$ and $V/I$ as a function of limb distance (obtained as explained in Sect.~\ref{sec:amplitude}) are shown in Fig.~\ref{fig:clvs}. 
We added estimations of the noise level, based on the standard deviation in the continuum per limb position and scaled it with the line depth-to-continuum ratio to account for the lower photon rate in the spectral lines. 
We display the noise level with red shaded areas in Fig.~\ref{fig:clvs}. 
The function $a(1-\mu^2)(\mu+b)^{-1}$ \citep[proposed by][]{Stenflo1997a} is used to fit the $Q/I$ CLV, where $a_{\mathrm{Fe}}=0.01\%$ and $b_{\mathrm{Fe}}=0.18$, and $a_{\mathrm{Sr}}=0.25\%$ and $b_{\mathrm{Sr}}=0.15$ for Fe~{\sc i} and Sr~{\sc i}, respectively. 
We note that we are not interested in a detailed modeling of the CLV of $Q/I$ in the depolarizing Fe~{\sc i} line. 
For this reason, we decided to consider the same fitting function, which for this line is exclusively used as a reference to locate the noise level.
Higher weight is given to small $\mu$ values, but the $\mu=0$ pixel is excluded. 
We point out that the Fe~{\sc i} line does not show any scattering polarization signal. 
Its linear polarization signal, which would only show signatures of the transverse Zeeman effect, can serve as zero reference. 
The CLV that is observed in the $Q/I$ of this line (see right $Q/I$ panel of Fig.~\ref{fig:clvs}) is a residual due to continuum depolarization  \citep[e.g.,][]{2001ASPC..236..141T,2001A&A...378..627F}.
In Sr~{\sc i}, the observed $Q/I$ amplitude very close to the limb is lower than expected by the fitted curve. 
Since the amplitude is especially sensitive to any shifts of the limb position with respect to the camera, a possible reason for its small value could be that over the long observing time and subsequent averaging the $Q/I$ amplitude may be smeared due to short-term (shorter than seconds) variations. 
Although we took special care that the long-term variability of the shifts is small, short-term variations cannot be excluded. Additionally, the analytical function has been obtained for the case of a plane-parallel atmosphere. Therefore, due to deviations from a plane-parallel geometry of the observed region close to the limb, the function could be unreliable at these positions.\par

We added the noise level to the fitted $Q/I$ CLV to reveal amplitude variations that are well above the statistical noise.
While in Fe~{\sc i,} the Stokes $Q/I$ amplitude, as a function of $\mu$, shows only small oscillations (which do not significantly exceed the noise level) around the fitted CLV curve, in Sr~{\sc i,} such oscillations are more pronounced.\par

Similarly to the $Q/I$ amplitudes, the $U/I$ amplitudes in Fe~{\sc i} remain for the majority of limb distances within the noise boundaries, which means that the statistical noise is also dominant here. 
 We find that the correction method based on slow modulation works remarkably well, as the absolute level of polarization in $U/I$ is very close to zero. 
Because of the noise domination, we exclude the presence of magnetic fields with a strong component perpendicular to the line-of-sight. Considering that the Land\'e factor of the Fe~{\sc i} line is higher than that of the Sr~{\sc i} line (1.5 compared to 1), the transverse Zeeman effect as a cause for the $U/I$ signals in Sr~{\sc i} can be ruled out with high certainty. 
Additionally, instrumental $V/I$$\rightarrow$$U/I$ cross-talk can be discarded, observing that $V/I$ is larger in Fe~{\sc i} and therefore this effect would be more pronounced in $U/I$ of this line. 
From the panels in Fig.~\ref{fig:obs_images}, it is also clear that 
$U/I$ in Sr~{\sc i} has a single-peaked profile and does not have a typical Zeeman shape.\par  
In $V/I$, most of the determined amplitudes are larger than the noise level, although small in magnitude, indicating weak magnetic activity.
The signals are larger in Fe~{\sc i} than in Sr~{\sc i}, as expected considering the larger Land\'e factor of the former.  
The $V/I$ profiles clearly vary along the slit (see $V/I$ panel in Fig.~\ref{fig:clvs}) and also the sign changes a few times, suggesting that unipolar magnetic structures responsible for the $V/I$ signals are not larger than a few arcsec in spatial direction. 
To evaluate the temporal stability of the $V/I$ profiles and therefore the temporal stability of the magnetic structures, we analyze the $V/I$ amplitudes for each of the 35 corrected Stokes images separately. Since the S/N of these images is not sufficient to fit all $V/I$ profiles (as done in Sect.~\ref{sec:amplitude}), we average over the red peak of the $V/I$ profiles in Fe~{\sc i}. The number of spectral pixels averaged is 15. We plot these averages as gray lines in Fig.~\ref{fig:clvs}, where each line represents one corrected image. 
We find that for $\mu < 0.2,$ these averaged values are less confined to the temporal mean (black line) than for $\mu > 0.2$. It is unclear if this occurs because of a low S/N or because of faster changing magnetic field configurations. 
For $\mu > 0.2$, evidence suggests that the magnetic structures are long-lived and the polarity within a pixel did not change dramatically during the observation. To estimate the apparent longitudinal magnetic flux density, equation (9.87) from \citet{LandiDeglInnocenti2004} was evaluated with the observed Sr~{\sc i} $V/I$ amplitude and assuming the collision parameters for this line given by \citet{Faurobert-Scholl1993}. 
This provides a value of about 1\,G. We stress that this estimate does not aim to give an accurate magnetic field value, but an order of magnitude with which the $V/I$ amplitude is compatible through the weak-field approximation. 
The estimated value emphasizes that the observed region was certainly very quiet and incorporates very weak structured magnetic fields. \par

With only very few exceptions, the pattern of the CLV of the $V/I$ amplitude in Fe~{\sc i} and in Sr~{\sc i} are very similar, indicating that both lines may interact with the same magnetic fields during formation. 
Remarkably, $V/I$ shows some correlation with $U/I$ only in Sr~{\sc i}. 
We analyze this aspect further in the next section. 

\subsection{Correlation of scattering polarization with circular polarization}
\label{sec:correlation}

\begin{figure}
\centering
\includegraphics[width=\hsize]{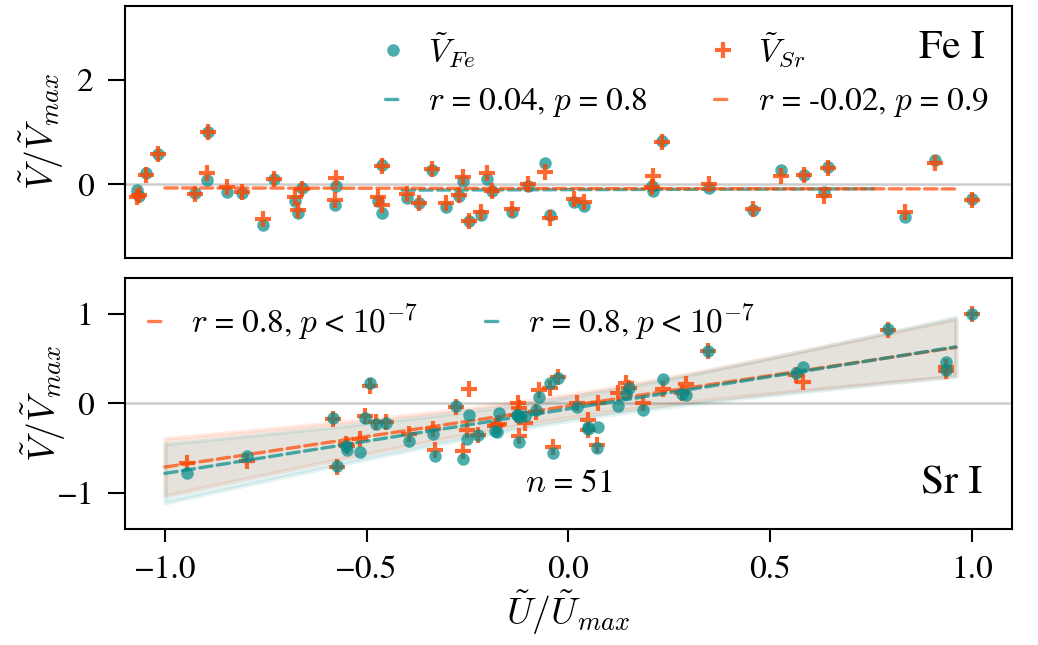}
\caption{
        Scatter and correlation of $\tilde{U}$ amplitudes in Fe~{\sc i} (\textit{top}) and Sr~{\sc i}  (\textit{bottom}) with circular polarization amplitudes $\tilde{V}$  of Fe~{\sc i} (circles) and Sr~{\sc i} (plus), respectively, scaled to their maximum values. The notation $\tilde{X}=X/I$ is used here. %Only limb distances $\mu > 0.2$ are considered. 
        The Pearson's correlation coefficient $r$ is determined together with the $p$-value by the linear regression fitted to the scattered data and displayed in the panels ($n$: total number of scattered points). The shaded area is the 95\% confidence interval for the fitted parameters.
}
\label{fig:correlation}
\end{figure}

In Fig.~\ref{fig:correlation}, we scatter the polarization amplitudes of $U/I$ with $V/I$ shown in Fig.~\ref{fig:clvs} for $\mu > 0.2$. 
To make a fair comparison between Fe~{\sc i} and Sr~{\sc i} that have different absolute amplitude values, we scale $U/I$ and $V/I$ of each spectral line to their respective absolute maximum value. 
We fit a linear regression to this data and calculate the Pearson correlation coefficient, $r$, and the corresponding $p$-value with the python package \textit{scipy}. 
In the top panel of Fig.~\ref{fig:correlation}, Fe~{\sc i}  $\tilde{U}/\tilde{U}_{max}$ is compared with $\tilde{V}/\tilde{V}_{max}$ of both Sr~{\sc i} and Fe~{\sc i} (where, for notational simplicity, we introduced $\tilde{X}=X/I$). 
Low $r$ values and $p\gg 0.05$ indicate that the null hypothesis that there is no correlation cannot be rejected.
Therefore, for Fe~{\sc i}, $\tilde{U}/\tilde{U}_{max}$ is independent of the Zeeman-induced signals in $V/I$.
This stands in contrast to what we find for $\tilde{U}/\tilde{U}_{max}$ in Sr~{\sc i}, which is shown in the bottom panel of Fig.~\ref{fig:correlation}. 
Here, the high $r$ values in combination with very low $p$ values suggest that the $U/I$ signals in Sr~{\sc i} correlate with the line-of-sight component of the magnetic field.

%-------------------------------------------------------------------   
\section{Discussion and conclusions}
\label{sec:discussion}
%------------------------------------------------------------------- 

At the IRSOL observatory, we performed spectropolarimetric measurements in the Sr~{\sc i} 4607.3\,\AA\ line applying a new technique based on slow modulation in front of the telescope. Combined with ZIMPOL, this technique provides high precision and enhanced accuracy, allowing for the absolute value of polarization to be well determined.
This allowed us to clearly detect, for the first time in the Sr~{\sc i} line, Stokes $U/I$ signals of different sign at different limb distances.
Our analysis of these measurements revealed a statistical significant correlation between such $U/I$ signals and the line-of-sight component of the magnetic field. The latter is inferred from the $V/I$ signals observed in the same Sr~{\sc i} line and in a nearby Fe~{\sc i} line. 
Instrumental causes for this correlation can be excluded. 
As the patterns of $V/I$ in Fe~{\sc i} and Sr~{\sc i} are very similar, we assume that both lines are exposed to similar magnetic fields during formation.
Nonetheless, the missing Stokes $U/I$ in the adjoining Fe~{\sc i} line indicates that the transverse Zeeman effect was too weak to be detected. 
Furthermore, at any limb distance, it is clear that the spectral shapes of the Sr~{\sc i} $U/I$ profiles are singly peaked.
Thus, the Zeeman effect induced by solar magnetic fields can also be excluded as a source for the Stokes $U/I$ signals in the Sr~{\sc i} line. 
The possibility that the Sr~{\sc i} $U/I$ signals are due to horizontal inhomogeneities that break the axial symmetry of the problem \citep[i.e., granulation, such as that found by][]{Zeuner2020} is ruled out by the long integration time of over an hour. This long integration time implies that the spatio-temporal resolution is too low to resolve granules, which have a lifetime of only several minutes \citep{Hirzberger1999}.

In conclusion, it is only Hanle rotation by deterministic (i.e., long-lived and structured on the resolution scale of the observation) magnetic fields that remains as the most plausible explanation for these $U/I$ signals in Sr~{\sc i}, which so far have never been detected in the photosphere with this level of clarity.
It is widely known that the Hanle effect only operates in the Doppler core of spectral lines, and since Sr~{\sc i} is a sharp resonance line without extended wings, the Hanle effect impacts the full profile.
This means that the Hanle effect in Sr~{\sc i} may not only serve as a diagnostic tool for turbulent magnetic fields in the photosphere, as demonstrated extensively by \citet{Stenflo1982, Faurobert-Scholl1995,TrujilloBueno2004, Shchukina2011a}. In addition, it has the potential to supplement and constrain Zeeman diagnostics for weak deterministic magnetic fields, especially in regions where a transverse Zeeman effect is too weak to be detected. \par
In the future, we plan to interpret our observations through detailed radiative transfer calculations. 
In the modeling process, it will be necessary to simultaneously include a turbulent magnetic field that is needed to reproduce the amplitude of $Q/I$ in Sr~{\sc i} together with its CLV, along with a deterministic field needed to interpret the signatures of Hanle rotation. 
Previously, \citet{LopezAriste2007} showed that in Sr~{\sc i} observations at THEMIS the Stokes $Q/I$ amplitude is correlated with Stokes $V/I$. These authors interpreted this finding as a Hanle effect by deterministic field on top of Hanle effect by turbulent magnetic field.
It is interesting to note that these magnetic fields survive for a substantial amount of time and can hardly be associated with the granulation, and also their approximate spatial scale is larger than a typical granule.
If these signals are found to be ubiquitous, they would challenge the view that the very weak, small-scale component of the magnetic field is purely turbulent and rapidly changing. 

Studying the temporal evolution of such long-lived, weak magnetic fields with the Hanle effect will shed light on the interaction of existing magnetic fields with a local dynamo. 
Such studies require, on the one hand, for an investigation of the co-existence of a structured, weak magnetic component with the turbulent component within 3D magneto-hydrodynamic simulations provided by codes such as MURaM \citep{2005A&A...429..335V} or CO5BOLD \citep{2012JCoPh.231..919F}; on the other hand, it also calls for the use of large solar telescopes to achieve a high S/N within a short amount of time, such as with the Daniel K. Inouye Solar Telescope on Maui, Hawaii, together with techniques to determine the absolute level of polarization, as will be made possible thanks to the planned European Solar Telescope. 

%-------------------------------------------------------------------   
   \begin{acknowledgements}
   We thank Daniel Gisler for clarifying technical discussions.
   IRSOL is supported by the Swiss Confederation (SEFRI), Canton Ticino, the city of Locarno and the local municipalities. 
   This project has received funding from the European's Horizon 2020 research and innovation programme under grant agreement no 824135. Measuring solar polarization with high absolute accuracy was proposed and developed with support of the ERC Advance Grant project HotMol ERC-2011-AdG-291659.
    L.B. and N.G. gratefully acknowledge the Swiss National Science Foundation for financial support through grant CRSII$5\_180238$.
    All the plots were done using the \textit{matplotlib} python package.
   \end{acknowledgements}
   
% BIBLIOGRAPHY
\bibliographystyle{bibtex/aa}
\bibliography{bibtex/bib}
%-------------------------------------------------------------------

\end{document}